\documentclass[10pt,letterpaper,twocolumn]{article} 

\usepackage{ol2}
\usepackage{hyperref}
\usepackage{amsmath}
\usepackage{epstopdf}

\newcommand{\fig}[1]{Fig.~\ref{#1}}
\newcommand{\eq}[1]{Eq.~(\ref{#1})}
\newcommand{\Fig}[1]{Figure~\ref{#1}}
\newcommand{\mf}[1]{\ensuremath{\mathcal{#1}}}		
\newcommand{\klr}[1]{\ensuremath{\left( #1 \right)}}  

\renewcommand{\_}[1]{\ensuremath{_\mathrm{#1}}}

\begin{document}

\twocolumn[ 

\title{Weak-signal conversion from 1550\,nm to 532\,nm with 84\% efficiency}


\author{Aiko Samblowski,$^{1}$ Christina E.~Vollmer,$^1$ Christoph Baune,$^1$ Jarom\'{i}r Fiur\'{a}\v{s}ek,$^2$ and \\ Roman Schnabel$^{1,*}$}

\address{
$^1$Institut f\"ur Gravitationsphysik, Leibniz Universit\"at Hannover and Max-Planck-Institut f\"ur Gravitationsphysik (Albert-Einstein-Institut), Callinstr.~38, 30167~Hannover, Germany\\
$^2$Department of Optics, Palack\'y University, 17.\ listopadu 12, 77146 Olomouc,
Czech Republic\\
$^*$Corresponding author: roman.schnabel@aei.mpg.de
}

\begin{abstract}
We report on the experimental frequency conversion of a dim, coherent continuous-wave light field from 1550\,nm to 532\,nm with an external photon-number conversion efficiency of (84.4\,$\pm$\,1.5)\%. We used sum-frequency generation, which was realized in a standing-wave cavity built around a periodically poled type\,I potassium titanyl phosphate (PPKTP) crystal, pumped by an intense field at 810\,nm. Our result is in full agreement with a numerical model. For optimized cavity coupler reflectivities it predicts a conversion efficiency of up to 93\% using the same PPKTP crystal.
\end{abstract}

\ocis{190.4223, 120.3940, 270.5585.}

] 

\noindent 
The efficient frequency conversion of photons is an important task in photonics to shift light fields into the visible wavelength regime where high quantum efficiency single-photon detectors are available \cite{Albota2004, VanDevender2007}. In principle an arbitrarily dim propagating light field can be up-converted to another frequency with an efficiency of 100\% preserving the photon number and its statistics \cite{Kumar1990}. 
The required energy needs to be supplied in terms of an intense pump field. The dim signal field and the pump field need to be overlapped inside a phase-matched nonlinear crystal, possibly supported by a cavity that is resonant for one or several of the wavelengths involved.

For \emph{intense} continuous-wave (cw) light we realized second-harmonic generation (SHG) with an external conversion efficiency of up to 95\% taking into account all optical losses \cite{Ast2011}. The  SHG process, however, is not suitable for the frequency up-conversion of dim fields, i.e. for the frequency up-conversion of arbitrarily weak signals.

In this Letter, we report on a highly efficient frequency conversion of a dim coherent light field from 1550\,nm to 532\,nm in the cw regime using (non-degenerate) sum-frequency generation. 
The overall conversion efficiency was measured to be $(84.4\pm 1.5)$\%, including all conversion and optical propagation losses.

Our experiment used a Nd:YAG laser that provided 2\,W cw output power at 1064\,nm. The entire field was directly sent to a second-harmonic-generation (SHG) cavity to generate up to 1.5\,W of second-harmonic light at 532\,nm. This beam was filtered by a mode-cleaning (MC) cavity to ensure a TEM\_{00} spatial mode profile and to suppress technical noise. The MC cavity finesse was $\mf{F}=560$, corresponding to a linewidth of 1.3\,MHz. The cavity length was controlled using the Pound-Drever-Hall (PDH) locking scheme \cite{Black2001} with a phase modulation at a sideband frequency of 29.5\,MHz.

To provide the signal field at 1550\,nm and the pump field at 810\,nm, a non-degenerate, doubly resonant optical parametric oscillator (OPO) was pumped with the filtered 532\,nm light. The OPO was built as a monolithic standing-wave non-linear cavity and was operated above threshold. The non-linear medium inside the OPO cavity was a periodically poled potassium titanyl phosphate (PPKTP) crystal. The phase matching for 532, 810 and 1550\,nm was achieved at a temperature of 68$^\circ$C, to which the crystal was stabilized actively. The length of the crystal was 8.9\,mm and the coatings were chosen to form a cavity with a finesse of $\mf{F}=100$ for the twin beams at 810\,nm and 1550\,nm. Hence, the linewidth of both modes was 91\,MHz. The radii of curvature of 8\,mm led to a waist size of 24\,$\mu$m for the green pump beam, that simply double-passed the crystal. The threshold power was about 70\,mW. The bright output fields were co-propagating and spatially separated with a dichroic beam splitter. 

\begin{figure}[tb]
\centerline{\includegraphics[width=\columnwidth]{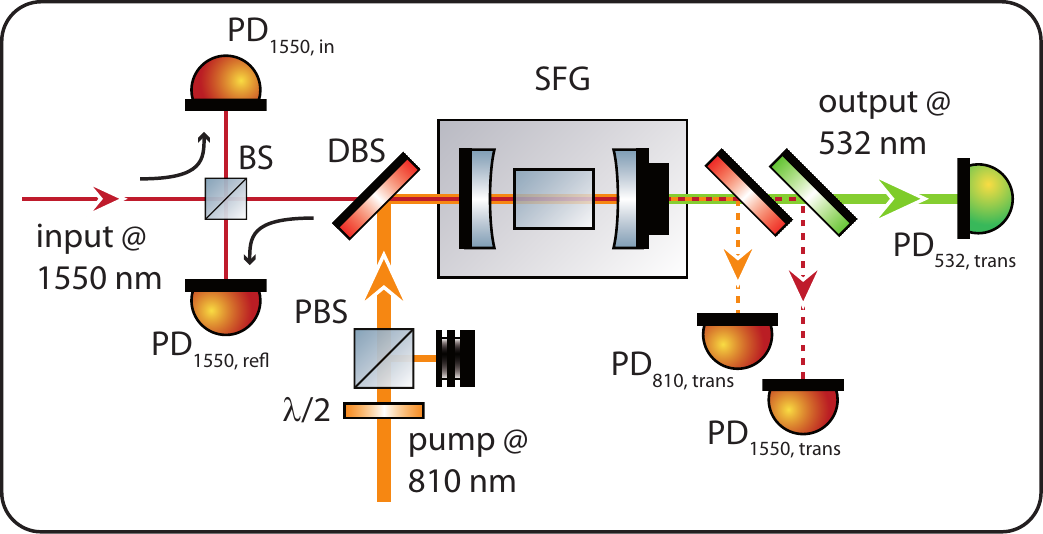}}
\caption{(Color online) Schematic of the core setup. SFG: sum-frequency generator; (P)BS: (polarizing) beam splitter; PD: photo diode; $\lambda/2$: waveplate for power variation. The generation of the intense pump field at 810\,nm and the generation of the weak signal field at 1550\,nm are not shown. Details are provided in the main text.}
\label{fig:setup}
\end{figure}

The pump field at 810\,nm with a tunable power of up to 190\,mW was mode-matched into the sum-frequency generator (SFG), see Fig.~\ref{fig:setup}. The SFG was built as a standing-wave two-mirror non-linear cavity containing another PPKTP crystal. The quasi phase matching for 810 and 1550\,nm was achieved at a temperature of 67$^\circ$C, to which the crystal was also stabilized actively. The length of the crystal was 9.3\,mm and the coatings were chosen to be $R>99.9\%$ on the right side and {$R=(96.5\pm0.5)\%$} on the left side for the signal and pump beam, see \fig{fig:setup} for illustration. The radii of curvature of 25\,mm and the air gaps of 17\,mm led to a waist size of about 50\,$\mu$m for the pump beam. The coatings for the converted light at 532\,nm are chosen to be $R<0.1\%$ on the right side and $R>99.9\%$ on the left side to ensure that the converted light leaves the SFG to the right. The length of the cavity was actively controlled with the PDH scheme, using a frequency modulation of the pump light at 24.5\,MHz. The losses of the system are mainly determined by the anti-reflective coatings of the crystal surface and by the absorption of the crystal itself. They can be combined in a total absorption of the crystal, that was measured to be {$\alpha\_{810}=0.46\%/$cm} and {$\alpha\_{1550}=0.19\%/$cm}. 

A 4\,mW signal beam at 1550\,nm (red) is sent through a 50/50 beam splitter (BS) to monitor the input power with the photo detector PD\_{1550, in}, see \fig{fig:setup}. The remaining 2\,mW incident power is overlapped with the pump beam at a dichroic beam splitter (DBS) and coupled into the SFG. Half of the reflected light is detected at PD\_{1550, refl}. In transmission of the SFG, the optical fields are separated by a set of DBSs and detected with the corresponding photo detectors PD\_{532, trans}, PD\_{810, trans} and PD\_{1550, trans}. All signals from the photo detectors are recorded with a data acquisition system (\emph{PCI-6133} from \verb|National Instruments|) and analyzed with PC software (\emph{Labview}).

To characterize the conversion efficiency, the light powers of the optical fields were measured with the photo detectors mentioned above, which were calibrated with power meters. The conversion efficiency was calculated as the ratio of the converted photons at 532\,nm and the initial photons at 1550\,nm 
\begin{align}\label{eq:eta_Power}
\eta &= \frac{\text{QE}_{1550}}{\text{QE}_{532}}\cdot\frac{n\_{532}}{n\_{1550}} = \gamma\cdot\frac{532\cdot P\_{532}}{1550 \cdot P\_{1550}}\,.
\end{align}
Here, $\gamma$ denotes a correction factor for the quantum efficiency (QE) of the photo detectors and the power meters.

To determine the correction factor $\gamma$, the depletion of the signal field at 1550\,nm was measured in reflection and in transmission of the cavity, respectively. 
When no pump light was coupled into the SFG, the light reflected by the cavity far from its resonance, corresponding to the total incident power, and the light transmitted by the cavity on resonance were measured and normalized to unity. 
The relative depletion 
\begin{align}\label{eq:eta_Ast}
\notag
\delta &= 1-\klr{\frac{P\_{refl}+P\_{trans}}{P\_{in}}}\\
&= 1-\Bigg(\frac{P\_{refl}}{P\_{refl,max}}+\underbrace{\frac{P\_{trans,max}}{P\_{refl,max}}}_{\kappa}\cdot\frac{P\_{trans}}{P\_{trans,max}}\Bigg)
\end{align}
depends on the normalized signals and on the ratio of the maximal transmitted and reflected power $\kappa$. 
In low loss systems, the relative depletion is a measure for the conversion efficiency \cite{Ast2011} and \eq{eq:eta_Ast} yields the same results as \eq{eq:eta_Power}. Due to the finesse of $\mf{F}_{1550}=150$ and to the total loss {$\alpha\_{1550}=0.19\%/$cm}, the two differ. 
However, the relative depletion provides additional data to determine the parameters of our theoretical model more precisely. In particular the correction factor $\gamma$ from \eq{eq:eta_Power} could be obtained accurately. Thus, both measurements are required to obtain an accurate value for the conversion efficiency.

To measure the conversion efficiency the pump power was varied. For every pump power we took time series of each photo detector over a couple of seconds and analyzed them in a \emph{Labview} script. The time series did not vary and the system was stable over hours.

\begin{figure}[t]
\centerline{\includegraphics[width=\columnwidth]{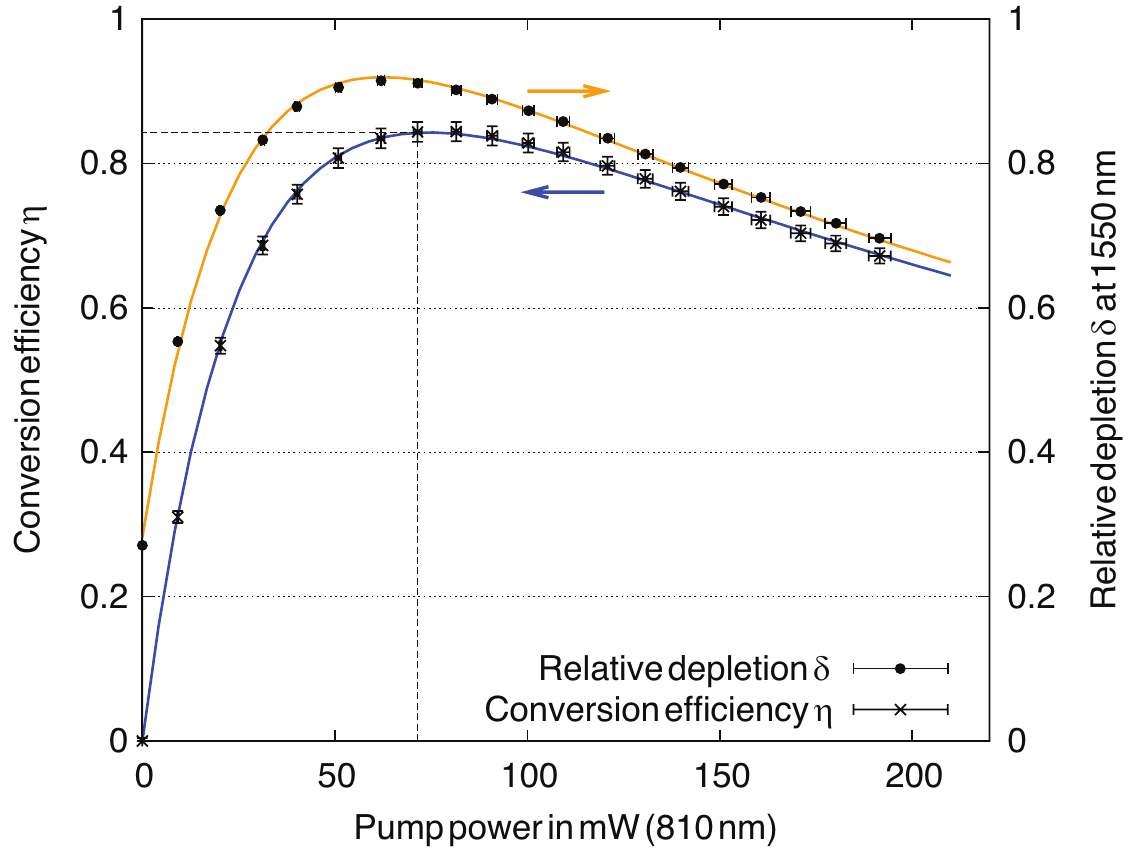}}
\caption{(Color online) Measurement results. The conversion efficiency (blue) and relative depletion (yellow) are shown over the pump power. The conversion efficiency reaches its maximum of (84.4\,$\pm$\,1.5)\% at 81.5\,mW pump power. The solid lines correspond to a numerical simulation of the system.}
\label{fig:theory}
\end{figure}

\Fig{fig:theory} shows the results of our measurement. The conversion efficiency (blue) and the relative depletion (yellow) are plotted against the pump power. The solid lines depict numerical simulations of our system, including mirror reflectivities, the non-linearity, losses and the phase mis\-matching as parameters. The simulations are in excellent agreement with our measurements and support the experimentally obtained conversion efficiency of (84.4\,$\pm$\,1.5)\%. As predicted for sum-frequency generation, energy is transferred back and forth between the interacting fields \cite{Kumar1990}. Hence, the conversion efficiency drops after reaching its maximum at a pump power of 81.5\,mW.

To determine the conversion efficiency of our up-conversion cavity a numerical time-domain simulation was used. 
Starting from an empty cavity, the field-strength evolutions for all three wavelengths involved were calculated in small steps along the propagation axis. If after many cavity round trips the field evolution reached its steady state the calculation stopped \cite{Lastzka2010}. 
The numerical simulation of conversion efficiency and relative depletion determined the calibration factor of the power meters to $\gamma = 1.0337\pm5\cdot 10^{-4}$, which was well within the specified error range of the power meters. 
Furthermore, our simulation showed that it should be possible to improve the conversion efficiency of our setup to about 93\% by reducing the mirror reflectivity at the signal wavelength from $R=96.5\%$ to $R=90\%$. 

Since sum-frequency generation maintains the quantum properties of a state \cite{Kumar1990}, our device will be suitable to reach a high fidelity if a non-classical state of light is used as an input field. Squeezed vacuum states with a noise suppression of more than 12\,dB have been demonstrated at a wavelength of 1550\,nm in the continuous-wave regime \cite{Mehmet2011}. In contrast, comparatively small squeezing factors were achieved at visible wavelengths so far \cite{Tsuchida1995}.
Assuming that the main optical loss is given by the imperfect conversion efficiency and assuming that the detection efficiency at 532\,nm will be the same as at 1550\,nm, 
non-degenerate sum frequency  generation might be a feasible approach for producing strongly squeezed states of light in the visible regime.\\

This work was supported by the Deutsche Forschungsgemeinschaft (DFG), Project No.\ SCHN 757/4-1, by the Centre for Quantum Engineering and Space-Time Research (QUEST) and by the International Max Planck Research School for Gravitational Wave Astronomy (IMPRS-GW). J.F. acknowledges support from the European Social Fund and MSMT under project No. EE2.3.20.0060.

\end{document}